\documentclass[]{aastex}
\usepackage{emulateapj5,onecolfloat,graphicx}

\def\eg{{\it e.g.}}
\def\etal{{\it et al.}}
\def\etc{{\it etc.}}
\def\ie{{\it i.e.}}

\def\BTii{BT08}

\long\def\Ignore#1{\relax}

\slugcomment{To appear in ApJ Letters}

\shorttitle{Relaxation in simulations}
\shortauthors{Sellwood}

\begin{document}

\twocolumn[
\title{Relaxation in $N$-body simulations of disk galaxies}

\author{J. A. Sellwood}
\affil{Department of Physics and Astronomy, Rutgers University, \\
    136 Frelinghuysen Road, Piscataway, NJ 08854}
\email{sellwood@physics.rutgers.edu}

\begin{abstract}
I use $N$-body simulations with two mass species of particles to
demonstrate that disk galaxy simulations are subject to collisional
relaxation at a higher rate than is widely assumed.  Relaxation
affects the vertical thickness of the disk most strongly, and drives
the velocity ellipsoid to a moderately flattened shape similar to that
observed for disk stars in the solar neighborhood.  The velocity
ellipsoid in simulations with small numbers of particles quickly
approaches this shape, but shot noise also dominates the in-plane
behavior.  Simulations with higher, but reachable, numbers of
particles relax slowly enough to be considered collisionless, allowing
the in-plane dispersions to rise due to spiral activity without
heating the vertical motions.  Relaxation may have affected many
previously published simulations of the formation and evolution of
galaxy disks.
\end{abstract} 

]

\section{Introduction}
Simulations over several decades have contributed a great deal to our
understanding of the complex dynamical behavior of stellar systems.
The numbers of particles employed in typical work has risen steadily
with the available computational power, but even today we are unable
to employ as many particles as there are stars in a galaxy, at least
on a routine basis.  Thus particles in simulations are typically fewer
in number and more massive than are stars in a galaxy, and therefore
the density distribution suffers from a higher level of shot noise
than expected in the stellar component of the system being mimicked.

\citet{Chan41} estimated the relaxation time, or the time required for
deflections by other stars to cause a star to lose all memory of its
original trajectory, to be
\begin{equation}
\tau_{\rm relax} = {v^3 \over 8\pi G^2\mu^2n \ln\Lambda},
\label{eq.relax}
\end{equation}
where $\mu$ is the mass of a star that moves at the typical speed $v$
and $n$ is the number of stars per unit volume.  The ratio $\Lambda
\simeq R_e/b_{\rm min}$, with $R_e$ being the half-mass radius of the
galaxy, and $b_{\rm min} \simeq 2G\mu/v^2$, the impact paramater below
which the approximation that deflections are small fails badly.  The
Coulomb logarithm, $\ln\Lambda$, indicates that every decade of
increase of the impact parameter makes an equal contribution to the
relaxation rate, because the number of stars rises with distance to
compensate for the diminishing influence of each individual encounter.
In simulations, one generally replaces $b_{\rm min}$ with the gravity
softening length, $\epsilon$, but distant encounters are unaffected by
softening and fully contribute to relaxation.  The reason for
softening the interparticle force is mostly to avoid the need for
shorter time steps during close approaches between particles
\citep[][hereafter \BTii, p123]{BT08}.

Following \BTii, we set $v^2 \simeq GN\mu/R_e$, $n \simeq 3N / (4\pi
R_e^3)$, and define a dynamical time $\tau_{\rm dyn} = R_e /v$.
With these substitutions, eq.~(\ref{eq.relax}) becomes
\begin{equation}
\tau_{\rm relax} \simeq {N \over 6 \ln N}\; \tau_{\rm dyn},
\label{eq.relax2}
\end{equation}
indicating that star-star encounters are utterly negligible for
galaxies.  The effect of a halo of particles whose masses are much
smaller than those of stars is further to lengthen the relaxation
time, since it increases $v$ without contibuting to the deflections.
Eq.~(\ref{eq.relax2}) also suggests that simulations with $N \ga
10^5$ should be safely collisionless, especially as the $\ln N$ term
in the denominator is an overestimate for the values of $\epsilon$
usually employed.

While scattering in disks is more rapid, see \S2, relaxation caused by
star-star encounters remains much too slow to determine the random
speeds of stars in the solar neighborhood.  The local distribution of
peculiar velocities \citep[\eg][]{Nord04, HNA09} has a flattened
tri-axial shape.  The dispersion in the direction of the Galactic
center, $\sigma_R$, is the largest component, with that in the orbital
direction, $\sigma_\phi$, being about 70\% as great as expected from
epicycle theory (\BTii, p170), while the dispersion, $\sigma_z$,
normal to the Galactic plane is some 60\% of $\sigma_R$, and is the
smallest.

In an important paper, \citet{Ida93} showed that the shape of the
velocity ellipsoid is consistent with scattering by dense mass clumps,
such as giant molecular clouds (GMCs).  Their predicted flattening
depends on the slope of the Galactic rotation curve, but were this
locally flat, they predict $\sigma_z \simeq 0.6\sigma_R$, as observed.
While GMCs readily redirect peculiar motions, they are less efficient
at increasing them; GMCs are believed to be insufficiently massive or
numerous \citep{Lace91, HF02} to create the high dispersion of the
oldest thin-disk stars.

Disk stars are also believed to be scattered at the Lindblad
resonances of transient spirals \citep{CS85}, which increase only the
in-plane dispersions.  The vertical dispersion should be unaffected
because the rapid vertical oscillation of a star is adiabatically
invariant at the slow rate at which it encounters the spiral density
variations \citep{Carl87}.  Thus spiral scattering is needed to
account for the high peculiar speeds of the older thin-disk stars
while massive gas clumps redirect the peculiar velocities to maintain
the shape of the velocity ellipsoid \citep{Sell13}.

Most simulations of galaxy disks do not include a separate population
of heavy particles, and therefore have no agent to redirect the random
motions generated by spiral waves into the vertical direction if they
were truly collisionless.  However, a few authors, notably
\citet{QHF93} and \citet{MD07}, have worried that isolated disks
thicken in simulations as spiral activity heats the in-plane motions.
Here I show that collisional relaxation is the likely origin of their
finding, and should be a concern for all simulations of disk galaxies.

\section{Relaxation in disks}
Formulae (\ref{eq.relax}) and (\ref{eq.relax2}) were derived assuming
a mass distribution that is roughly uniform in 3D.  The relaxation
rate in disks differs for at least four reasons, as were mostly
pointed out by \citet{Rybi72}.

First, because disks are rotationally supported, stars pass each other
at speeds that are a small fraction of the circular orbit speed, say
$\beta v$, with $\beta \simeq 0.1$.  The predominantly slow encouters
mean that the typical deflection rate is boosted by the factor
$\beta^{-1}$, and the time needed for the random impulses to amount
to a peculiar speed of $\beta v$ is shorter by a factor $\beta^3$.

Second, eq.~(\ref{eq.relax}) results from an integration of
mean-square deflections over all impact parameters.  The 3D
integration volume element becomes an element of area in 2D, and
therefore the impact parameter enters with one power less.  Thus the
Coulomb logarithm is replaced by the factor $(b_{\rm min}^{-1} -
b_{\rm max}^{-1})$, indicating that scattering is dominated by close
encounters.  Distant encounters are negligible in disks because the
number of stars does not rise rapidly enough with distance to
compensate for the diminished force each exerts.

Real galaxy disks are neither razor thin, nor spherical.  In this
case, the spherical dependence applies at ranges up to the typical
disk thickness, $z_0$, beyond which the contribution to scattering
drops quickly.  Thus we should replace $R_e$ in the Coulomb logarithm
by $z_0$, which slightly reduces the relaxation rate.

Third, the local number density of stars is higher so that $N \sim \pi
R_e^2z_0n$, which increases the density to be used in
eq.~(\ref{eq.relax}) by the factor $R_e / z_0$.  This factor causes
another order of magnitude increase in the relaxation rate.

Combining these three considerations yields a relaxation time in disks
that is shorter than in 3D distributions by the factor
\begin{equation}
\beta^3\left({z_0 \over R_e} \right){\ln\left(R_e / b_{\rm min}\right)
  \over \ln\left(z_0 / b_{\rm min}\right)},
\label{eq.2D}
\end{equation}
which is typically a few $\times 10^{-4}$!  It should be noted that
the cube of $\beta$ arises from the time needed for scattering to
produce a peculiar velocity $\beta v$, \ie\ equal to that of a typical
disk star; the relaxation {\it rate\/} is increased over the 3D rate
by only a single factor of $\beta$, but together with the higher
density, the increase is still $\sim 100$-fold.

The fourth factor in real disks is the existence of GMCs, whose role
in determining the shape of velocity ellipsoid was described in the
introduction.

\section{Simulations}
Here I report a few simulations to test these theoretical predictions;
a fuller study will be presented elsewhere.

Since simulations of all realistic disks heat due to instabilities,
simple measurements of the heating rate will not determine the
relaxation rate directly.  Therefore, following \citet{Hohl73}, I
employ separate particle species having different masses, since energy
exchange between particles of differing mass is a reliable indicator
of relaxation.

The smooth, half-mass Mestel disk (\BTii, p99) was proved by
\citet{Toom81} to have no global instabilities, although
\citet{Sell12} found subtle instabilities, caused by non-linear
effects, that gave rise to spiral activity in large-$N$ models.  It
seems unlikely that 3D motion would alter the unusual stability
properties of this disk, which make it more attractive for this test
than other more generic disks that could host global linear
instabilities.

\subsection{Setting up the disk}
In order to construct an equilibrium model, I start from the
two-integral DF for the razor-thin disk \citep{Zang76, Toom77}:
$f_{\rm Z}(E,L_z) \propto L_z^qe^{-E/\sigma_R^2}$, where $E$ is a
particle's specific energy and $L_z$ its specific $z$-angular
momentum.  The parameter $q$ determines the radial velocity dispersion
$\sigma_R = V_0(1+q)^{-1/2}$, with $V_0$ being the circular orbital
speed at all radii.  The value of Toomre's local stability parameter
for a razor-thin Mestel disk is
\begin{equation}
Q \equiv {\sigma_R \over \sigma_{R,{\rm min}}} = {2^{3/2}\pi \over 3.36 f(1+q)^{1/2}},
\label{Qsingle}
\end{equation}
where $f$ is the active mass fraction in the disk; both $\sigma_R$ and
$Q$ are independent of radius.  In order obtain a disk with $\beta \la
0.15$, which must also have $Q>1$, I adopt $f = 0.25$, with the
remaining mass in a rigid halo.

I apply inner and outer tapers to limit the radial extent of the disk:
$f_0(E,L_z) = f_{\rm Z}[1+(L_i/L_z)^4]^{-1}[1+(L_z/L_o)^6]^{-1}$,
where $L_i$ and $L_o$ are the central angular momentum values of the
inner and outer tapers respectively.  I choose $L_o = 15L_i$, and
further restrict the extent of the disk by eliminating all particles
whose orbits would take them beyond $20R_i$, where $R_i = L_i/V_0$ is
the central radius of the inner taper.  With these tapers $R_e \simeq
8R_i$.

I thicken the disk by giving it the Gaussian vertical density profile
$ \rho(R,z) = \Sigma(R)\exp(-z^2/2z_0^2)/ (2\pi z_0), $ with
$\Sigma(R)$ being the vertically integrated surface mass density at
radius $R$.  I estimate the equilibrium vertical velocity dispersion
at each $z$-height by integrating the 1D Jeans equation (\BTii,
eq.~4.271) in the numerically-determined potential.  I adopt an
unrealistically small value $z_0 = 0.05R_i$, that is independent of
$R$, in order to reveal the effects of relaxation clearly.

The active mass gives rise to a weaker central attraction in the
midplane than that which should arise from a razor-thin, infinite,
full-mass Mestel disk.  In order to maintain equilibrium, I add a
rigid central attraction to the grid-determined forces from the
particles in the disk at each step.  This unchanging, spherically
symmetric, rigid central attraction is pre-tabulated from differencing
$-V_0^2/R$ from the grid-determined attraction of a smooth density
created from the thickened, tapered disk.

\subsection{Numerical details}
I determine the gravitational field of the particles using the 3D
polar grid described in \cite{SV97}.  The grid has $200$ rings, $256$
spokes, and $375$ vertical planes, and I adopt a cubic spline
softening rule that yields the full attraction of a point mass at
distances $\geq 2\epsilon$.  I choose $R_i = 5$ grid units, the grid
planes are $0.02R_i$ apart, $\epsilon = 0.025R_i$, a basic timestep of
$0.025R_i/V_0$, and advance particles at radii $R>2R_i$ at intervals
that double in duration from this radius and again with every factor 2
increase in $R$.  The duration of the simulations was $\sim 20$ full
rotations of the disk at $R_e$, or $\sim 3\;$Gyr when scaled to the
Milky Way.

In order to create two populations of particles with unequal masses, I
employ every particle selected from the DF twice, placing the two
particles each at separately chosen random azimuths, and make one 9
times more massive than the other.  The total masses of both
populations are set to yield the desired total disk mass ($f=0.25$ in
eq.~\ref{Qsingle}) and a combined $Q=1.5$, which corresponds to
$\sigma_R \simeq 0.14V_0$ and $\sigma_\phi \simeq 0.1V_0$, \ie\ in the
ratio expected for a flat rotation curve (\BTii).  Thus $\sigma_{\rm
  total}/V_0 = \beta \simeq 0.17$.

\begin{figure}[t]
\begin{center}
\includegraphics[width=\hsize,angle=270]{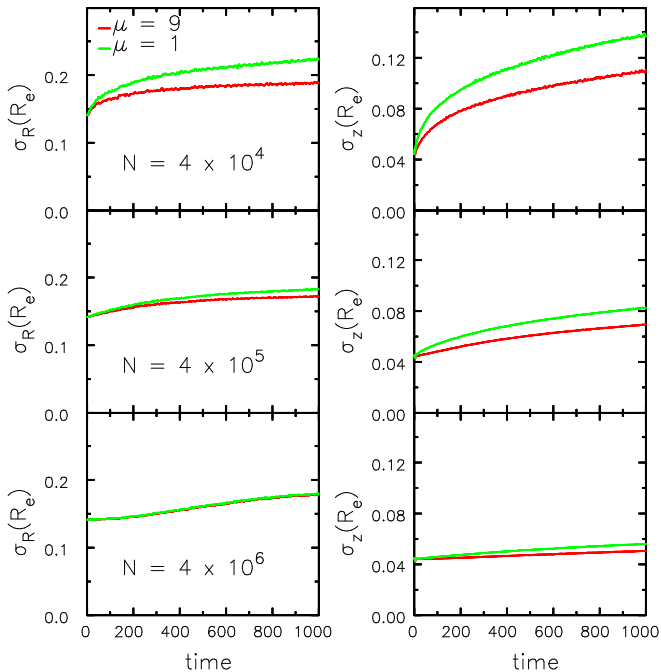}
\end{center}
\caption{Measurements over a broad swath of the disk around the
  half-mass radius of the time evolution of the average $\sigma_R$
  (left) and $\sigma_z$ (right).  The top row shows results with $N=2
  \times 10^4$ in each species and the particle number was increased
  10-fold from row to row.  The red (green) curves apply to the heavy
  (light) particles respectively.  Velocities are in units of $V_0$,
  distances in units of $R_i$, and the rotation period at $R_e$ is
  $\sim 50$ in these time units, where $G = V_0 = R_i = 1$.}
\label{fig.results1}
\end{figure}

\begin{figure}[t]
\begin{center}
\includegraphics[width=\hsize,angle=270]{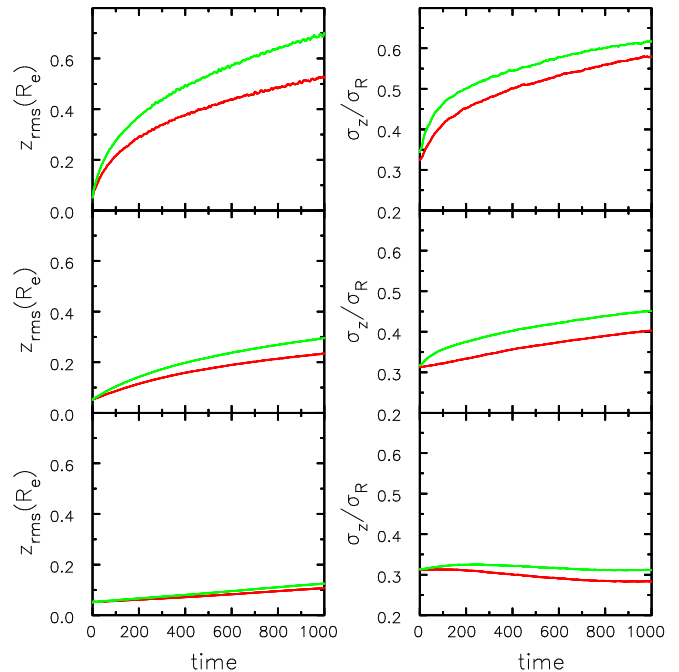}
\end{center}
\caption{Continuation of Fig.~\ref{fig.results1}: rms $z$-thickness
  (left) and the ratio $\sigma_z/\sigma_R$ (right).}
\label{fig.results2}
\end{figure}

\subsection{Results}
Figure~\ref{fig.results1} shows the evolution of $\sigma_R$ and
$\sigma_z$, averaged over a broad radial range centered on $R_e$,
measured for both the heavy (red) and light (green) particles in three
simulations.  Each row is from a separate simulation, differing in the
number of particles, as indicated in the left-hand panels.  Since the
simulations start with randomly placed particles, the initial behavior
is dominated by swing-amplified shot noise \citep{TK91}, and I stop
the calculations after $\sim 20$ rotations at $R_e$, while spiral
activity is on-going.

The radial velocity dispersion rises due to spiral activity, which has
lower initial amplitude as the number of particles rises.  The
vertical dispersion and rms $z$-thickness (Figure~\ref{fig.results2})
of each population rises rapidly in the smallest simulation, but are
almost constant in the largest, demonstrating that the models were in
initial equilibrium.  However, the tendency for mass segregation,
unmistakable in the top row, is still visible in the bottom row.

The different heating rates for the two mass species is clear evidence
for relaxation, as is also the rapid rise in the vertical motions of
both populations as the number of particles is decreased.  Notice
also, from the top right panel of Figure~\ref{fig.results2} that the
velocity ellipsoid shapes of both populations rapidly become rounder.
The larger experiments, on the other hand, manifest a slower energy
equipartition rate and also support the theoretical prediction that
spirals do not heat vertical motions.

The rate at which $v_\perp^2$ rises due to encounters, one step before
arriving at eq.~(\ref{eq.relax}), is
\begin{equation}
{d v_\perp^2 \over dt} = {8\pi G^2\mu^2n \ln\Lambda \over v},
\end{equation}
In the simulations, $\mu \simeq M_d/N_h$ with $N_h$ being the number
of heavy particles, $\Lambda = z_0/\epsilon$, $v = \beta V_0$, and $n
\sim N_h / (\pi R_e^2z_0)$.  We set $M_d = fM_{\rm tot}$, and $V_0^2 =
GM_{\rm tot}/(R_e)$, and assume that the vertical velocity dispersion,
$\sigma_z$, is rising due to relaxation.  Thus we expect
\begin{equation}
{d \sigma_z^2 \over dt} = {8f^2V_0^3 \ln(z_0/\epsilon) \over N_h \beta
  z_0} \sim {40 \over N_h},
\end{equation}
if $V_0=1$, $f=0.25$, $z_0=R_i/20 = 2\epsilon$, and $\beta \simeq 0.17$.

I estimate the initial value of $d\sigma_z^2/dt \sim 1.1 \times
10^{-5}$ from the light particles in the intermediate simulation.
This is to be compared with $40/N_h = 2 \times 10^{-4}$, since $N_h=
2\times 10^5$.  Thus the predicted scattering rate is about 20 times
that observed.  This discrepancy appears to be due to some additional
grid smoothing in the simulation.  Relaxation is even slower when
softening and grid cell sizes are increased, and also when fewer
sectoral harmonics, $m$, contribute to the grid-determined forces.  In
Figs.~\ref{fig.results1} \& \ref{fig.results2}, the force calculation
discarded contributions to the forces from $m > m_{\rm max}=32$, and
the relaxation rate in further simulations drops by about 30\% for
each factor 2 reduction in $m_{\rm max}$.  Applying a cut-off in the
expansion of the density distribution effectively smooths each
particle in the azimuthal direction, weakening the attraction between
disk particles and slowing the relaxation rate.  Since relaxation is
inhibited by the extra smoothing in grid methods, I have deliberately
employed a finer grid and more sectoral harmonics than is normally
required in order to highlight relaxation.

A dependence on grid smoothing is physically reasonable, but contrasts
with the finding of \citet{HB90} that the relaxation rate does not
depend on the simulation method.  Note that they were testing
spherical models, where relaxation is dominated by density
fluctuations on large scales and is little unaffected by short-range
smoothing, which is the reason they found the rate of relaxation to be
the same in every valid numerical method.  However, relaxation in
disks is dominated by close encounters (\S2), making the rate much
more dependent on the degree of smoothing.  Therefore, it seems
reasonable to expect a higher relaxation rate for the same $N$ in
direct methods, \eg\ tree codes, than in grid-based methods.

\section{Discussion and conclusions}
Both the theoretical argument in \S2 and the numerical results in \S3
confirm that two-body scattering in disks is much more rapid than that
expected in a spheroidal model with the same $N$.  This is because
disk particles pass each other at speeds much lower than the orbital
speed, and because the density of particles in a disk is higher than
were the same number spherically distributed.  Furthermore, unlike in
3D systems, relaxation is dominated by the particles within a few disk
thicknesses, and the cumulative effect of distant encounters is less
important.

The experiments also confirm that spirals, which heat the in-plane
motions (left panel of Fig.~\ref{fig.results1}), do not cause even
extremely thin disks to thicken, when relaxation is slow enough
(bottom row).  With fewer particles, relaxation causes the shape of
the velocity ellipsoid of both species to evolve rapidly to roughly
the shape observed in the solar neighborhood (top right panel of
Fig.~\ref{fig.results2}).

\citet{MD07} tried to investigate why the disks in their simulations
thickened.  They found that thickening was inhibited in parallel
experiments in which the position and velocities of each disk particle
were rotated through a random angle after each time step.  Having
suppressed collective spiral responses by this stratagem, they
concluded that thickening was due to spiral heating.  However,
two-body relaxation arises from the time-integrated perturbing forces
as particles pass, which must also have been suppressed in this test.

Possible relaxation needs to be taken into account when interpreting
results from simulations in a wide variety of contexts.  For example,
the number of star particles genrally employed in disk formation
simulations is in the range of a few $\sim 10^5$, as reported in the
code comparison by \citet{Scan12}.  Simulations of star-forming disks
could be particularly severely affected by relaxation, since new stars
formed from cold gas will have the smallest $\beta$.  \citet{Hous11}
compare the thickening of their simulated disk with SDSS data, but it
is unclear what this test shows because relaxation probably caused
some thickening in their model.  Relaxation may have contributed to
disk thickening in the simulation of dwarf disk galaxy formation by
\citet{Gove10}, which employed $\sim 5 \times 10^5$ particles.  Other
areas where relaxation may have affected the conclusions include
studies of the survival of thin disks \citep[\eg][]{Robe06, Most10}
and thickness variations during radial migration in disks
\citep[\eg][]{Minc11, Rosk12, Bird13}.  In particular, the actions of
particles, when calculated exactly by \citet{SSS12}, were conserved
only on average in their simulations with $N \ga 10^6$; it seems
likely that the inexact conservation of this quantity could have been
due to slow relaxation.

Since disk thickening is most strongly affected by relaxation, the
outcome of a simulation will depend on the number of particles
employed.  On the one hand, the shape of the velocity ellipsoid is
determined by relaxation in simulations with modest $N$.  The heavy
particles scatter each other somewhat as GMCs scatter stars in
galaxies so that, paradoxically, low-quality simulations get the right
shape of the velocity ellipsoid for the wrong reason!  But in
small-$N$ simulations, the in-plane dynamics is dominated by the
collective responses to shot noise.

With large $N$, on the other hand, collisionless in-plane dynamics is
more faithfully represented and coupling of spiral heating to the
vertical motion is weak.  But to mimic the evolution of the velocity
ellipsoid, one would need to include a population of extra-heavy
particles to redirect the in-plane motions.  Note that these heavy
particles would also affect the in-plane dynamics \citep{TK91, DVH12}.

Thus any simulation of an isolated stellar disk that thickens
probably does so through 2-body relaxation.  If gas is included,
clumps of gas particles may behave as scattering centers, as do the
GMCs in real disks, but with a smaller mass ratio to the star
particles.  Simulations that mimic the full hierarchical evolution
will have other sources of vertical heating, such as in-falling dwarf
galaxies and sub-halos, \etc \ However, segregation of star particles
of different masses would remain a valid diagnostic of relaxation in
all these contexts.

\section*{Acknowledgments}
The author thanks Tad Pryor and Michael Solway for helpful
conversations, Victor Debattista for comments on a draft, and the
Editor for advice.  This work was supported in part by NSF grant
AST-1108977.

\end{document}